\documentclass[11pt,twoside]{article}

\usepackage{asp2006}
\usepackage{epsf}
\usepackage{psfig}
\usepackage{lscape}

\begin{document}
\title{Bulge formation by the coalescence of giant
clumps in primordial disk galaxies}   
\author{Bruce G. Elmegreen}   
\affil{IBM Research Division, Yorktown Heights, NY 10598, USA, bge@us.ibm.com}    

\begin{abstract} The observations and evolution of clumpy, high-redshift
galaxies are reviewed. Models suggest that the clumps form by
gravitational instabilities in a gas-rich disk, interact with each
other gravitationally, and then merge in the center where they form a
bulge. The model requires smooth gas accretion during galaxy growth.
\end{abstract}


\section{Introduction: Secular Evolution of Classical Bulges}   

There is growing evidence for secular bulge evolution, even for
classical bulges. The classical bulge fraction (i.e., with high Sersic
index, $n$) decreases out to $z\sim1$ (Sargent et al. 2007); bulge and
disk colors correlate with each other, suggesting that classical bulges
and disks coevolve (Balcells \& Dominguez-Palmero 2008); the Milky Way
old thick disk and bulge have similar metallicities, suggesting
co-evolution (Mel\'endez et al. 2008); the central disk mass
concentration (which is essentially the bulge to disk ratio) increases
over time at $z\sim2$ (Genzel et al. 2008); and young disks have clumpy
structure and young bulges are similar to disk clumps (Elmegreen et al.
2008). In addition, simulations suggest that disks with high gas
fractions and high turbulent speeds form massive clumps that move to
the center and make a ``classical'' bulge (Noguchi 1999; Immeli et al.
2004; Bournaud, Elmegreen, Elmegreen 2007, Elmegreen, Bournaud, \&
Elmegreen 2008ab). At the same time, $\Lambda$CDM cosmological
simulations have not reproduced the observed small value of
$M(bulge)/M(disk)$ (e.g., Graham \& Worley 2008), suggesting there were
few major mergers during disk formation (Weinzirl et al. 2008).

\section{Chain Galaxies and Clump Clusters}

We have been observing the properties of star-forming clumps in
galaxies in the Hubble Space Telescope Ultra Deep Field (UDF), and in
the GEMS and GOODS fields. The UDF shows clumpy, star-bursting galaxies
out to $z\sim6$, where the Lyman $\alpha$ Hydrogen line begins to shift
out of the reddest band in the Advanced Camera for Surveys. A catalog
of galaxy types is in Elmegreen et al. (2005a). Of greatest interest
are the extremely clumpy galaxies, the chains and the clump-clusters
(Fig. 1), which comprise 31\% of all 1003 UDF galaxies larger than 10
pixels in diameter. In comparison, spirals comprise another 31\%,
ellipticals 13\%, and likely interacting galaxies, the double-clump and
tadpoles, 25\%. The clumpy types are somewhat uniformly distributed
over redshift out to $z\sim5$, while the spirals and ellipticals
concentrate within $z\sim1.5$ because of their intrinsically red colors
(Elmegreen et al. 2007). These fractions emphasize the importance of
the clumpy phase of galaxy evolution. Their co-moving density (chains +
clump clusters) is about $10^{-3}$ Mpc$^{-3}$ for $z>2$, with an
increase to $\sim3\times10^{-3}$ Mpc$^{-3}$ for $z<1$ (Elmegreen et al.
2007). In comparison, the co-moving densities of spirals and
ellipticals in the UDF are $\sim4\times10^{-3}$ and
$\sim1\times10^{-3}$ Mpc$^{-3}$, respectively. It is conceivable,
considering the short life of the clumpy phase, that all spiral
galaxies go through this phase before they settle down to a smooth disk
with a bulge in the center.

\begin{figure}\plotone{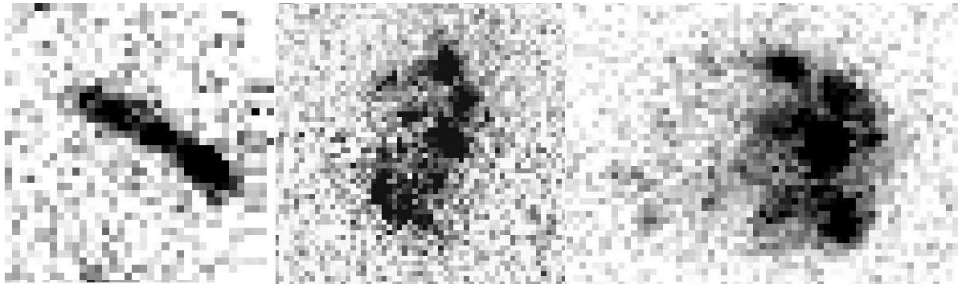}
\caption{Chain (left) and clump cluster galaxies from the background
field of the Tadpole galaxy, which was one of the first deep images
taken with the HST ACS camera (From Elmegreen, Elmegreen \& Hirst
(2004).}\end{figure}

We believe there is a link between chain galaxies, which are linear
alignments of clumps, and clump clusters, which are round or irregular
conglomerates of clumps. The galaxy and clump properties are the same
in these two types (Elmegreen, Elmegreen, \& Hirst 2004), and their
combined distribution on a histogram of the ratio of axes
(width/length) is nearly flat with a drop off below $\sim0.2$ and above
$\sim0.8$ (Elmegreen \& Elmegreen 2005). A disk viewed in random
projection has a flat distribution on such a plot, with a fall-off at
low ratio from the relative disk thickness. There should be no fall-off
at high ratio for a uniform disk, but for a clumpy disk, there is
always some irregularity at the edge because of the discrete clump
positions. This irregularity fits the clumpy galaxy distribution for
the observed number of 5-10 clumps per galaxy (Elmegreen \& Elmegreen
2005). This result implies that chain galaxies are edge-on clump
clusters.

The clump positions in a chain galaxy are also consistent with their
being edge-on projections of clump positions in disk galaxies. The
clumps are not uniformly spaced, for example, as might be the case for
condensations in a filament. Figure 2 shows relative clump positions in
all of the UDF chains, measured relative to the end clumps. The curve
is a model for clumps that are distributed randomly in azimuth around a
disk, and exponentially in radius on average with two scale lengths (as
observed for clump clusters -- Elmegreen et al. 2005b). The similarity
between the curve and the histogram suggests again that chain galaxies
are edge-on clump clusters.

\begin{figure}
\psfig{file=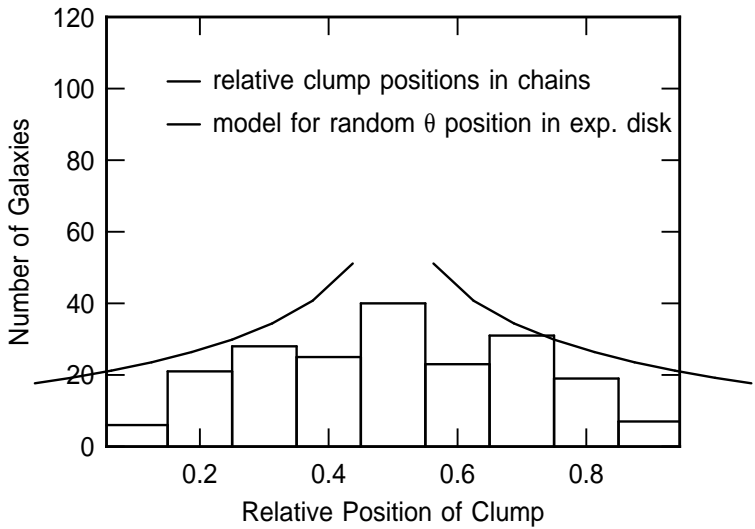}
\caption{Average clump distribution along the length of a chain galaxy
is plotted as a histogram. The curves show a model distribution for an
edge-on disk with an exponential radial profile having two scale
lengths for the average clump position and a random angular
distribution around the center.}\end{figure}

It is important to establish this projection link between chain
galaxies and clump clusters because the chain galaxies have their
clumps highly confined to the mean line connecting them. They deviate
in position from this mean line by an average distance of only
$\sim100$ pc. This is much smaller than the average thickness of the
galaxies themselves, which is $z_0\sim900$ pc for scale height $z_0$ in
the functional fit ${\rm sech} z/z_0$.  We infer from this that the
clumps in both chains and clump clusters are confined to a disk
midplane, and therefore that they formed there. The alternative
possibility, that the clumps entered the galaxy from outside (as in a
hierarchical build up scenario) seems untenable with their observed
midplane positions.  For in-situ formation, the clumps are most likely
giant star-forming regions that formed in a gas disk. This makes them
similar to star complexes in modern galaxies, although the high
redshift clumps are more massive by a factor of $\sim100$ than today's
complexes (Elmegreen et al. 2009).

\section{Bulges in Clumpy Galaxies}

The clumpy large galaxies in the UDF (diameters $>10$ pixels) were
examined on both ACS images in optical bands and NICMOS images in
near-IR bands. In 30\% of the chains and 50\% of the clump clusters,
bulges or bulge-like objects can be seen in the NICMOS images even if
they are not present or easily recognized in the ACS images.  In the
other cases, there are no obvious bulges, although there is usually
some NICMOS emission associated with either the brightest clumps or the
underlying disk.

ACS + NICMOS measurements of bulges-like objects were made in 47 clump
clusters and 27 chains. Bulges were also measured in 131 spiral
galaxies.  ACS measurements of clumps were made in 184 clump clusters,
112 chains, and 118 spirals, yielding 898, 406, and 845 clumps
respectively. We fit the clump SEDs to Bruzual \& Charlot (2003) models
with free parameters for clump age, star formation exponential decay
time, and extinction. We included Madau (1995) intergalactic hydrogen
absorption, as well as Calzetti et al. (2000) and Leitherer et al.
(2002) extinction. The metallicity was assumed to be 0.4 solar because
of measurements like this in one case (Bournaud et al. 2008). The
metallicity dependence for mass and age are small anyway. Then we get
the clump mass from its brightness, star formation history, and
extinction. A complete discussion is in Elmegreen et al. (2009).

\begin{figure}
\plotone{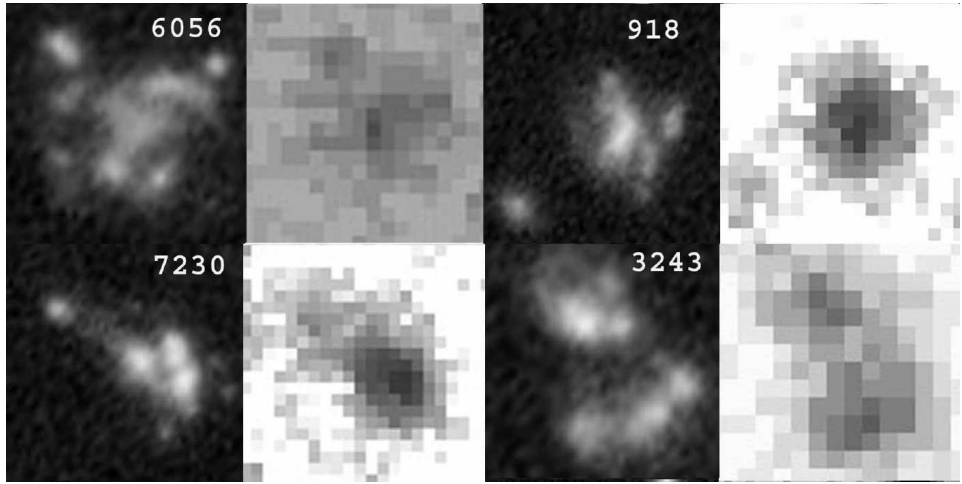} \caption{Four clump cluster galaxies are
shown, each with an ACS image on the left (a grayscale version of the
Skywalker color image, containing all of the ACS passbands) and a
NICMOS version on the right, with $3\times$ lower resolution. The top
two galaxies have NICMOS concentrations in the center that could be
bulges, while the bottom two galaxies have diffuse NICMOS emission only
from the star forming regions, which are not centralized. }\end{figure}

Figure 3 shows two clumpy galaxies with bulges (top) and two without
bulges (bottom). UDF numbers are indicated. Each galaxy is represented
by two panels, the left showing a grayscale version of the color image
from
Skywalker\footnote{http://www.aip.de/groups/galaxies/sw/udf/index.php},
at the full ACS resolution, and the right showing the NICMOS image at
$3\times$ worse resolution. There are dense central objects in the
NICMOS images for the top two galaxies, and only diffuse IR emission
following the star formation in the bottom two galaxies.

The measurements in this survey indicated that the bulges in clumpy
galaxies are a lot like the clumps in both age and mass, whereas in
spiral galaxies, the bulges are significantly more massive than the
clumps, and also much older. Thus the bulges look recently formed in
the clumpy cases, and similar enough to the clumps to lead us to think
that they formed from the clumps. The relative youth of the bulges in
clumpy galaxies is consistent with bulge morphology because often the
bulge-like objects are slightly off center and irregular for the clumpy
cases, and the surrounding disks are not generally exponential along
any radial cut. Spiral galaxies have more regular radial profiles and
more centralized clumps.

\section{Star Formation in Clumpy Galaxies}

The regular spacing of clumps in tidal arms, their midplane positions
in chains, the similarity between their sizes and the disk thicknesses
in chains, and their universally high masses all point to clump
formation by gaseous gravitational instabilities. The clumps contain a
stellar mass $M\sim10^7-10^8\;M_\odot$, and may contain gas too for the
youngest cases, perhaps even $10^8-10^9\;M_\odot$ of gas. The clump
size is typically $3\times$ the local star complex size (2000 pc
compared to 600 pc -- Efremov 1995), and the clump mass is typically
$100\times$ the local star-complex mass (which is
$10^4-10^5\;M_\odot$).

We can use these size and mass scalings to determine the implications
for the interstellar medium. The cloud size for gravitational
instabilities is $L_{\rm Jeans}\sim\sigma^2/G\Sigma$, and the cloud
mass is $M_{\rm Jeans}\sim\sigma^4/G^2\Sigma$.  The scale-up then
implies at high redshift, $\sigma\sim5\times$ the local $\sigma$, which
means $\sigma\sim40$ km s$^{-1}$. This is the dispersion observed for
the ionized component of the gas in these galaxies (Forster Schreiber
et al. 2006; Genzel et al. 2006; Weiner et al. 2006). Also at high
redshift, $\Sigma\sim10\times$ the local $\Sigma$, which means
$\Sigma\sim100\;M_\odot$ pc$^{-2}$ in the gas (this corresponds to
$1\times10^{22}$ H cm$^{-2}$ for the neutral ISM and has not been
observed yet unless it causes the damped Lyman $\alpha$ absorption).

Evidently, clump clusters and chains are forming the inner disks and
bulges of today's spiral galaxies. The process of star formation is the
same as it is locally, but at higher velocity dispersion and gas column
density, and most likely at higher gas fraction too, as the disk forms
roundish clumps instead of swing amplified spirals.  The roundness
implies that the instability and energy dissipation is faster than a
shear time.

\section{Models of Bulge Formation by the Coalescence of Disk Clumps}

These observations have led us to re-investigate an old model by
Noguchi (1999) and Immeli (2004ab) in which a massively unstable disk
fragments into a few giant clumps that interact and migrate to the
center to make a bulge. We chose parameters typical of the clumpy
galaxies we see (e.g., mass, size) and assumed a high gas fraction
(50\% in typical cases) and a high initial velocity dispersion.  The
initial disk was uniform in density profile out to a sharp edge. It
breaks into $M_{\rm Jeans}$ clumps quickly, within a fraction of a
rotation, and the clumps form stars, interact gravitational with each
other and with the rest of the galaxy, and migrate to the center where
they coalesce. The result is a bulge surrounded by a double-exponential
disk.

\begin{figure}
\plotone{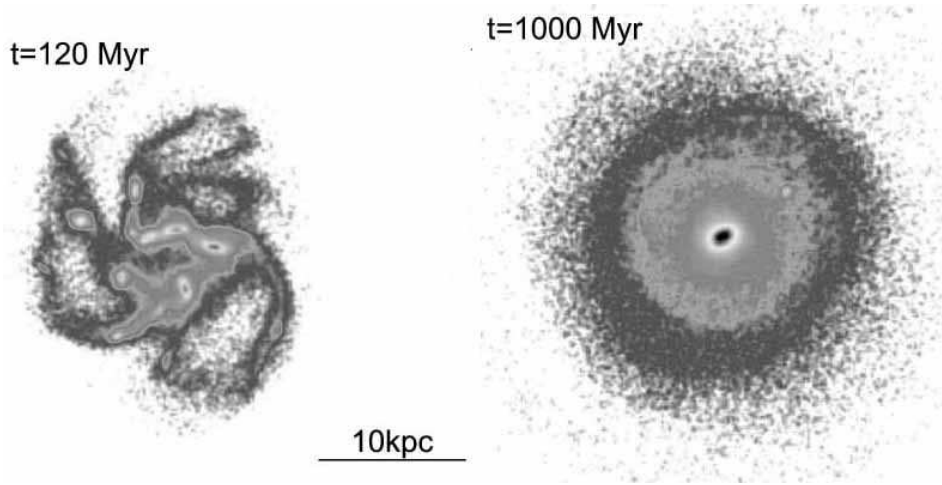} \caption{Model showing the early and late
stages of a typical simulation of a clump cluster galaxy, from BEE07.
Characteristic of the model is the high initial gas fraction in the
disk (50\%) and a high gas turbulent speed. The initial disk is uniform
with a sharp edge and half the size of the final disk. Clumps form
quickly with the Jeans mass, interact, and move to the center to make a
bulge. The final disk has a double-exponential radial profile, with the
outer part coming from clump debris.}\end{figure}

Figure 4 shows the basic model in 6 timesteps for a standard case. The
clumps form quickly, interact with each other, and migrate to the
center to make a bulge. Many different cases with about the same result
were discussed in Bournaud, et al. (2007). The bulge that forms is
slowly rotating and has a Sersic $n=4$ profile like a classical bulge
(Elmegreen et al. 2008b). If we put point particle tracers in the
center of each clump, representing intermediate mass black holes formed
by main sequence star coalescence in the dense cluster cores (e.g.,
Ebisuzaki et al. 2001), then these black holes migrate to the center
with the clumps, and there they can merge into a supermassive black
hole. The correlation between black hole mass and bulge velocity
dispersion is reproduced in this fashion, to within a factor of $\sim3$
(Elmegreen et al. 2008a).

There are several interesting results about clumps and their
coalescence.  First, feedback does not destroy the clumps because they
are too tightly bound. The high velocity dispersion of the average ISM
makes the clumps have a high dispersion too, and this makes them hard
to break apart by supernovae. Second, the swirling clumps in the center
mix the dark matter there just before they coalesce, and this can
decrease the density in any dark matter cusp by about a factor of 3.
Third, an existing bulge can tidally destroy the smaller disk clumps
and prevent them from accreting in the center. In this way, the bulge
formation process might be self-limiting.

\section{Conclusions}

There is a new galaxy morphology at $z>1$: disks with massive clumps.
The clumps appear to form in these disks by gravitational
instabilities. This conclusion follows from the similarity between the
clump sizes and the disk thicknesses, from the midplane positions, and
from the boundedness (relatively long lives) of the clumps. Clump
masses range from $10^7$ to $10^8$, and up to $10^9\;M_\odot$ in some
cases. The large mass is most likely the result of a large turbulent
speed of 25 to 50 km s$^{-1}$, and a large gas mass column density of
$\sim100\;M_\odot$ pc$^{-2}$. Dispersal and merging of the clumps
builds an exponential disk and a bulge.

Clumpy disks are observed to at least $z\sim5$ where they appear to
dominate spiral galaxies. This distribution suggests that spirals
evolve from relatively isolated clumpy disks and at least some of the
classical bulges in those spirals form by internal processes. The model
requires rapid gas assembly, however, not hierarchical merging of
pre-existing star-rich galaxies as in some hierarchical build-up
models.

There may be a way for the model to explain the Hubble type too: late
Hubble types could have a low gas velocity dispersion, in which case
the clumps would be relatively low-mass, they would not interact much,
and they would be easily destroyed by feedback.  The clumps would not
form a bulge in that case. Early Hubble types could have a high gas
dispersion, and this would make their clumps massive, strongly
interacting, somewhat impervious to star formation feedback, and
quickly brought to the center where they would merge to make a bulge.


\acknowledgements Most of this research was done in collaboration with
Debra Meloy Elmegreen, Frederic Bournaud, and numerous undergraduate
students at Vassar College.


\end{document}